\documentclass[checkin,prc,printnumbers,superscriptaddress,floatfix]{revtex4}

\usepackage{amsfonts}
\usepackage{graphicx}
\usepackage{graphicx,epsfig,latexsym,amssymb}
\usepackage{multirow,amsmath,array,booktabs}
\usepackage{dcolumn}
\usepackage[section]{placeins}
\usepackage{bm}

\begin{document}

\title{Relativistic effects of spin and pseudospin symmetries}
\author{Shou-Wan Chen}
\affiliation{School of Physics and Material Science, Anhui University, Hefei 230039,
People's Republic of China}
\author{Jian-You Guo}
\email[E-mail:]{jianyou@ahu.edu.cn}
\affiliation{School of Physics and Material Science, Anhui University, Hefei 230039,
People's Republic of China}

\begin{abstract}
Dirac Hamiltonian is scaled in the atomic units $\hbar =m=1$, which allows
us to take the non-relativistic limit by setting the Compton wavelength $%
\lambda \rightarrow 0 $. The evolutions of the spin and pseudospin
symmetries towards the non-relativistic limit are investigated by solving
the Dirac equation with the parameter $\lambda$. With $\lambda$
transformation from the original Compton wavelength to $0$, the spin
splittings decrease monotonously in all spin doublets, and the pseudospin
splittings increase in several pseudospin doublets, no change, or even
reduce in several other pseudospin doublets. The various energy splitting
behaviors of both the spin and pseudospin doublets with $\lambda$ are well
explained by the perturbation calculations of Dirac Hamiltonian in the
present units. It indicates that the origin of spin symmetry is entirely due
to the relativistic effect, while the origin of pseudospin symmetry cannot
be uniquely attributed to the relativistic effect.
\end{abstract}

\maketitle

It is well known that the spin and pseudospin symmetries play critical role
in the shell structure and its evolution. The introduction of spin-orbit
potential made the single-particle shell model can well explain the
experimentally observed existence of magic numbers for nuclei close to the
valley of $\beta $-stability~\cite{Haxel49,Mayer49}. To understand the near
degeneracy observed in heavy nuclei between two single-particle states with
the quantum numbers ($n-1,l+2,j=l+3/2$) and ($n,l,j=l+1/2$), the pseudospin
symmetry (PSS) was introduced by defining the pseudospin doublets ($\tilde{n}%
=n-1,\tilde{l}=l+1,j=\tilde{l}\pm 1/2$) \cite{Hecht69,Arima69}, which has
explained numerous phenomena in nuclear structure including deformation \cite%
{Bohr82}, superdeformation \cite{Dudek87}, identical bands \cite{Nazar90},
and magnetic moment \cite{Trolt94}. Because of these successes, there have
been comprehensive efforts to understand their origins as well as the
breaking mechanisms. For the spin symmetry (SS), the spin-orbit potential
can be obtained naturally from the solutions of Dirac equation. Thus, the SS
can be regarded as a relativistic symmetry. For the PSS, its origin has not
been fully clarified until now. It is worth reviewing some of the major
progresses in understanding the underlying mechanism of PSS. In Ref.\cite%
{Bahri92}, a helicity unitary transformation of a non-relativistic
single-particle Hamiltonian was introduced to discuss the PSS in the
non-relativistic harmonic oscillator. The particular condition between the
coefficients of spin-orbit and orbit-orbit terms was indicated in the
corresponding non-relativistic single particle Hamiltonian for the
requirement of PSS. The same kind of unitary transformation was considered
in Ref.\cite{Blokh95}, where the application of the helicity operator to the
non-relativistic single-particle wavefunction maps the normal state ($l,s$)
onto the pseudo-state ($\tilde{l},\tilde{s}$), while keeping all other
global symmetries. A substantial progress was achieved in Ref.\cite{Ginoc97}%
, where the relativistic feature of PSS was recognized. The pseudo-orbital
angular momentum $\tilde{l}$ is nothing but the orbital angular momentum of
the lower component of Dirac spinor, and the equality in magnitude but
difference in sign of the scalar potential $S$ and vector potential $V$ was
suggested as the exact PSS limit. Meng et al. showed that exact PSS occurs
in the Dirac equation when the sum of the scalar $S$ and vector $V$
potentials is equal to a constant \cite{Meng98}. Unfortunately, the exact
PSS cannot be met in real nuclei, much effort has been devoted to the cause
of splitting. In Ref. \cite{Alber01,Alber02,Lisboa10}, it was pointed out
that the observed pseudospin splitting arises from a cancellation of the
several energy components, and the PSS in nuclei has a dynamical character.
A similar conclusion was reached in Refs.~\cite{Marco01,Marco08}. In
addition, it was noted that, unlike the spin symmetry, the pseudospin
breaking cannot be treated as a perturbation of pseudospin-symmetric
Hamiltonian \cite{Liang11}. The non-perturbation nature of PSS has also been
indicated in Ref.\cite{Gonoc11}. Regardless of these pioneering studies, the
origins of the spin and pseudospin symmetries have not been fully understood
in the relativistic framework. Recently, we have checked the PSS by use of
the similarity renormalization group and shown explicitly the relativistic
origin of this symmetry \cite{Guo2012}. However, the dependence of the
quality of PSS on the relativistic effect has not been checked until now. In
this paper, we study the evolution of the spin and pseudospin symmetries
from the relativistic to the non-relativistic to explore the relativistic
relevance of this symmetries.

Dirac equation of a particle of mass $m$ in external scalar $S$ and vector $%
V $ potentials is given by
\begin{equation}
H=c\vec{\alpha}\cdot \vec{p}+\beta \left( mc^{2}+S\right) +V,
\label{Hamiltonian}
\end{equation}%
where $\vec{\alpha}$ and $\beta $ are the usual Dirac matrices. For a
spherical system, the Dirac spinor $\psi $ has the form
\begin{equation}
\psi =\frac{1}{r}\left(
\begin{array}{c}
iG_{n\kappa }\left( r\right) \phi _{\kappa m_{j}}(\vartheta ,\varphi ) \\
F_{n\kappa }\left( r\right) \vec{\sigma}\cdot \hat{r}\phi _{\kappa
m_{j}}(\vartheta ,\varphi )%
\end{array}%
\right) ,  \label{wavefunction}
\end{equation}%
where $n$ is the radial quantum number, and $m_{j}$ is the projection of
angular momentum on the third axis. $\kappa =\pm (j+1/2)$ with $-$ for
aligned spin ($s_{1/2}$, $p_{3/2}$, etc.), and $+$ for unaligned spin ($%
p_{1/2}$, $d_{3/2}$,etc.). Splitting off the angular part and leaving the
radial functions satisfy the following equation
\begin{equation}
\left(
\begin{array}{cc}
mc^{2}+\Sigma (r) & -c\frac{d}{dr}+\frac{c\kappa }{r} \\
c\frac{d}{dr}+\frac{c\kappa }{r} & -mc^{2}+\Delta (r)%
\end{array}%
\right) \left(
\begin{array}{c}
G\left( r\right) \\
F\left( r\right)%
\end{array}%
\right) =\varepsilon \left(
\begin{array}{c}
G\left( r\right) \\
F\left( r\right)%
\end{array}%
\right) ,  \label{Diraceq}
\end{equation}%
where $\Sigma (r)=V(r)+S(r)$ and $\Delta (r)=V(r)-S(r)$. Based on Eq.(\ref%
{Diraceq}), a lot of work has been carried out to check the origins of the
spin and pseudospin symmetries \cite%
{Ginoc97,Alber01,Alber02,Lisboa10,Liang11,Gonoc11}. Although they are
recognized as the symmetries of Dirac Hamiltonian, it is still not very
clear the important role of relativistic effect. In order to explore the
relativistic effects of this symmetries, the atomic units $\hbar =m=1$ are
adopted instead of the conventional relativistic units $\hbar =c=1$ in the
present system. For simplicity, the operator $H$ is measured in unit of the
rest mass, $mc^{2}$. Then the equation (\ref{Diraceq}) is presented as
\begin{equation}
\left(
\begin{array}{cc}
1+\lambda ^{2}\Sigma & \lambda \left( -\frac{d}{dr}+\frac{\kappa }{r}\right)
\\
\lambda \left( \frac{d}{dr}+\frac{\kappa }{r}\right) & -1+\lambda ^{2}\Delta%
\end{array}%
\right) \left(
\begin{array}{c}
G\left( r\right) \\
F\left( r\right)%
\end{array}%
\right) =\varepsilon \left(
\begin{array}{c}
G\left( r\right) \\
F\left( r\right)%
\end{array}%
\right) ,  \label{ALDiraceq}
\end{equation}%
where the Compton wavelength $\lambda =\hbar /mc=1/c$. In this units, the
result in the non-relativistic limit can be obtained in a very simple,
intuitive, and straightforward manner by taking the speed of light $%
c\rightarrow \infty $ or the Compton wavelength $\lambda \rightarrow 0$,
which is not possible in the latter units since $c=1$.

In order to investigate the evolution from the relativistic to the
non-relativistic, $\lambda $ is regarded as a parameter and the original
Compton wavelength $\lambda =\hbar /mc$ is labelled as $\lambda _{0}$. The
relativistic result corresponds to the solution of Eq.(\ref{ALDiraceq}) with
$\lambda =\lambda _{0}$. The result in the non-relativistic limit can be
obtained from Eq.(\ref{ALDiraceq}) by setting $\lambda \rightarrow 0$. Thus,
the evolution from the relativistic to the non-relativistic can be checked
by transforming $\lambda $ from $\lambda _{0}$ to $0$. Then, the
relativistic effects of the spin and pseudospin symmetries can be
investigated by extracting the energy splittings between the spin or
pseudospin doublets, and this symmetries develop toward the non-relativistic
limit can be checked, and vice versa.

In order to make this clear, we have solved Eq.(\ref{ALDiraceq}) for a
Woods-Saxon type potential for $\Sigma (r)$ and $\Delta (r)$, i.e., $\Sigma
(r)=\Sigma _{0}f(a_{\Sigma },r_{\Sigma },r)$ and $\Delta (r)=\Delta
_{0}f(a_{\Delta },r_{\Delta },r)$ with
\begin{equation}
f(a_{0},r_{0},r)=\frac{1}{1+\exp \left( \frac{r-r_{0}}{a_{0}}\right) }\text{.%
}  \label{Potential}
\end{equation}%
The corresponding parameters are determined by fitting the energy spectra
from the RMF calculations for $^{208}$Pb (to see Ref.~\cite{Guo053}). The
energy spectra of Eq.(\ref{ALDiraceq}) are calculated by expansion in
harmonic oscillator basis.

The single particle energy varying with the parameter $\lambda $ is
displayed in Fig.1, where it can be seen that the energy decreases
monotonously with $\lambda$ decreasing for all the levels available. The
trend of energy with $\lambda $ is towards the direction of non-relativistic
limit. With the decreasing of $\lambda $, the calculation is closer to the
non-relativistic result. When $\lambda $ is reduced to $\lambda /\lambda
_{0}=0.1$, the solution of Eq.(\ref{ALDiraceq}) is almost same as the
non-relativistic result. Furthermore, for the different single-particle
states, the sensitivity of energy to $\lambda $ is different. For the spin
unaligned states, the decreasing of energy is faster than that for the spin
aligned states, which leads to the energy splittings of the spin doublets
reduce with $\lambda $ decreasing. When $\lambda $ is reduced to $\lambda
/\lambda _{0}=0.1$, the spin-orbit splittings almost disappear for all the
spin doublets. These indicate that the spin symmetry becomes better as $%
\lambda $ decreases, and the spin symmetry breaking is entirely due to the
relativistic effect.

To better understand the preceding claim, the energies in several $\lambda $
values are listed in Table I for all single particle levels. For comparison,
Table I does also display the data of the non-relativistic calculations (the
last column), which are obtained by solving the Schr\"{o}dinger equation $%
H\psi (r)=E\psi (r)$ with $H=-\frac{\hbar ^{2}}{2m}\left( \frac{d^{2}}{dr^{2}%
}-\frac{l(l+1)}{r^{2}}\right) +\Sigma (r)$. From Table I, it can be seen
that the relativistic spin-orbit splitting ($\lambda =\lambda _{0}$) is
considerably large. This splitting decreases with the decreasing of $\lambda
$. When $\lambda /\lambda _{0}=0.001$, the energy of the spin unaligned
state in conjunction with that of the spin aligned state degenerates to the
non-relativistic result. These indicate that Eq.(\ref{ALDiraceq}) reproduces
well the process of development from the relativistic to the
non-relativistic, and both results of the relativistic and non-relativistic
can be obtained well from Eq.(\ref{ALDiraceq}) with an appropriate value of $%
\lambda $. Hence, the relativistic effects of the spin and pseudospin
symmetries can be checked from the solutions of Dirac equation with the
parameter $\lambda $.

In order to recognize clearly the relativistic effect of spin symmetry, the
energy splittings of spin doublets varying with $\lambda $ are plotted in
Fig.2, where it is shown that the energy splittings decrease monotonously
with $\lambda $ reducing for all the spin partners. When $\lambda $ is
reduced to $\lambda/\lambda_{0}=0.1$, the energy splittings of all the spin
doublets are almost reduced to zero. The detailed observation shows that the
energy splittings are more sensitive to $\lambda $ for the states with
higher orbital angular momentum in the same radial quantum number. For the
states with the same orbital angular momentum, the energy splittings appear
crosses in the different radial quantum number. These reflect that the
relativistic sensitivity is different for the states with different quantum
numbers. When $\lambda $ is reduced to zero, the spin-orbit splittings
disappear for all the spin partners, the non-relativistic results are
obtained in excellent agreement with those from the solutions of Schr\"{o}%
dinger equation. Namely, the spin-orbit splitting arises completely from the
relativistic effect, and can be treated as a perturbation of spin-symmetric
Hamiltonian as indicated in Ref.\cite{Liang11}.

Different from the spin symmetry, the relativistic origin of pseudospin
symmetry is more complicated. In Fig.3, we display the energy splittings of
pseudospin doublets varying with the parameter $\lambda $. From there, it
can be observed that the energy splittings increase significantly with $%
\lambda $ decreasing for the pseudospin partners ($2g_{9/2}$,$1i_{11/2}$), ($%
2f_{7/2}$,$1h_{9/2}$), and ($3p_{3/2}$,$2f_{5/2}$). Especially for ($2g_{9/2}
$,$1i_{11/2}$), the increasing of energy splitting is very obvious. For the
doublets ($2d_{5/2}$,$1g_{7/2}$), the increasing of energy splitting with $%
\lambda $ decreasing is relatively small. When $\lambda /\lambda _{0}$
decreases below than 0.6, the energy splitting goes toward a stable value.
The same phenomenon also appears in the doublet ($3s_{1/2}$,$2d_{3/2}$).
However for the pseudospin doublets ($2p_{3/2}$,$1f_{5/2}$) and ($2s_{1/2}$,$%
1d_{3/2}$), an opposite evolution of energy splitting with $\lambda $ is
disclosed. It shows the origin of pseudospin symmetry is more complicated
than that of spin symmetry. The pseudospin splitting cannot be attributed
uniquely to the relativistic effect. The quantum number of single-particle
states and the shape of potential make important influence on this symmetry.

In order to better understand the relativistic effects of the spin and
pseudospin symmetries, we expand perturbatively the Dirac Hamiltonian in Eq.(%
\ref{ALDiraceq}) to analyze the effects of each higher order term on the
energy splitting behaviors of both spin and pseudospin doublets. Following
Ref.~\cite{Guo2012}, for Dirac particle, the expanded Hamiltonian up to the
order $1/m^{3}$ is
\begin{eqnarray}
H &=&\Sigma \left( r\right) +\frac{p^{2}}{2m}-\lambda ^{2}\frac{1}{2m^{2}}%
\left( Sp^{2}-S^{\prime }\frac{d}{dr}\right) -\lambda ^{2}\frac{\kappa }{r}%
\frac{\Delta ^{\prime }}{4m^{2}}+\lambda ^{4}\frac{S}{2m^{3}}\left(
Sp^{2}-2S^{\prime }\frac{d}{dr}\right) +\lambda ^{4}\frac{\kappa }{r}\frac{%
S\Delta ^{\prime }}{2m^{3}}  \notag \\
&&+\lambda ^{2}\frac{\Sigma ^{\prime \prime }}{8m^{2}}-\lambda ^{2}\frac{%
p^{4}}{8m^{3}}-\lambda ^{4}\frac{{\Sigma ^{\prime }}^{2}-2\Sigma ^{\prime
}\Delta ^{\prime }+4S\Sigma ^{\prime \prime }}{16m^{3}},  \label{Schlike}
\end{eqnarray}%
where $p^{2}=-\frac{d^{2}}{dr^{2}}+\frac{\kappa (\kappa +1)}{r^{2}}$. Based
on the same considerations as Ref.~\cite{Guo2012}, $H$ is decomposed into
the eight components: $\Sigma \left( r\right) +\frac{p^{2}}{2m}$, $-\lambda
^{2}\frac{1}{2m^{2}}\left( Sp^{2}-S^{\prime }\frac{d}{dr}\right) $, $%
-\lambda ^{2}\frac{\kappa }{r}\frac{\Delta ^{\prime }}{4m^{2}}$, $+\lambda
^{4}\frac{S}{2m^{3}}\left( Sp^{2}-2S^{\prime }\frac{d}{dr}\right) $, $%
+\lambda ^{4}\frac{\kappa }{r}\frac{S\Delta ^{\prime }}{2m^{3}}$, $+\lambda
^{2}\frac{\Sigma ^{\prime \prime }}{8m^{2}}$, $-\lambda ^{2}\frac{p^{4}}{%
8m^{3}}$, $-\lambda ^{4}\frac{{\Sigma ^{\prime }}^{2}-2\Sigma ^{\prime
}\Delta ^{\prime }+4S\Sigma ^{\prime \prime }}{16m^{3}}$, which are
respectively labelled as $O_{1},O_{2},\cdots ,O_{8}$. $O_{1}$ corresponds to
the Hamiltonian in the non-relativistic limit, i.e., the Schr\"{o}dinger
part of $H$. $O_{2}(O_{4})$ is the dynamical term relating to the order $%
1/m^{2}(1/m^{3})$. $O_{3}(O_{5})$ is the spin-orbit coupling corresponding
to the order $1/m^{2}(1/m^{3})$. The eigenvalues of $H$ are calculated with
the fully same $\Sigma (r)$ and $\Delta (r)$ as that in calculating the
exact solutions of Eq.(\ref{ALDiraceq}).

For recognizing the relativistic effect of SS, we analyze the reason why the
energies of the spin unaligned states decrease faster than those of the spin
aligned states when $\lambda $ decreases. As an illustrated example, we
display the energy splittings of every component $O_{i}(i=1,2,\cdots ,5)$
varying with $\lambda $ for the spin doublets $(1p_{1/2},1p_{3/2})$ and $%
(1g_{7/2},1g_{9/2})$ in Fig.\ref{spinsplitting}, where we neglect the
results of $O_{6},O_{7},$and $O_{8}$ because their contributions to the
energy splitting are minor and do not influence on the total energy
splitting behavior with $\lambda $. From Fig.\ref{spinsplitting}, it can be
seen that the contributions of all the $O_{i}(i=2,3,4,5)$ to the energy
splittings between the spin unaligned states and the spin aligned states are
positive, and the positive energy splittings decrease with $\lambda $
decreasing. It is for this reason that the energies of the spin unaligned
states decrease faster than those of the spin aligned states with $\lambda $
decreasing. Compared with $O_{3}$ (the spin-orbit coupling corresponding to
the order $1/m^{2}$), and the contributions of $O_{2}$, $O_{4}$, and $O_{5}$
to the spin energy splittings are relatively minor. The total energy
splittings are dominated by the contribution of $O_{3}$ when $\lambda $ is
sufficiently small. This means that, as the relativistic effect becomes
weak, the spin splittings are almost entirely due to the spin-orbit
coupling. For the different spin partners, the energy splitting behaviors
with $\lambda $ are same except for the extent of splittings, as displayed
in Fig.\ref{spinsplitting} for the spin doublets $(1p_{1/2},1p_{3/2})$ and $%
(1g_{7/2},1g_{9/2})$. These indicate that the spin symmetry origins
completely from the relativistic effect, and possesses the perturbation
attribute claimed in Ref.\cite{Liang11}. To understand the relativistic
effect of PSS, we analyze the cause of the various energy splitting
behaviors of pseudospin doublets. In Fig.\ref{pseudosplitting}, we show the
energy splittings of each component $O_{i}(i=1,2,\cdots ,5)$ varying with $%
\lambda $ for the pseudospin partners $(2s_{1/2},1d_{3/2})$ and $%
(2f_{7/2},1h_{9/2})$. From there, it can be seen that the pseudospin energy
splittings caused by the Schr\"{o}dinger part of $H$ are dominated. This
splittings are reduced by the contribution of spin-orbit coupling, and added
by the contribution of dynamical terms. For the pseudospin partner $%
(2s_{1/2},1d_{3/2})$, with the decreasing of $\lambda $, the contribution of
the pseudospin breaking (the dynamical terms) declines faster than that of
the pseudospin improvement (the spin-orbit coupling), which results in
better PSS when $\lambda $ decreases. However for the $(2f_{7/2},1h_{9/2})$,
the energy splittings caused by the pseudospin breaking varying with $%
\lambda $ are relatively slower than that by the spin-orbit coupling, which
leads to the PSS becomes worse with $\lambda $ decreasing. These cause the
different energy splitting behaviors of pseudospin doublets with $\lambda $.
Hence, the pseudospin splitting can not be regarded as a perturbation in
agreement with the claim in Ref.\cite{Liang11}.

In addition to the energy splittings associated with the relativistic
effects, the wave function splittings between the (pseudo)spin doublets are
also associated with the relativistic effects. An illustrated example is
displayed in Fig.\ref{spinwf}, where the upper component of Dirac spinor for
the states $1g_{7/2,9/2}$ is depicted in several $\lambda $ values. From Fig.%
\ref{spinwf}, it can be seen that the wavefunction splitting of spin doublet
is obvious for a relativistic particle ($\lambda /\lambda _{0}=1$). With the
development towards the non-relativistic direction (to reduce $\lambda $),
the wavefunction splitting of spin doublet decreases, which is in agreement
with the case of level splitting. For the pseudospin symmetry, the lower
component of Dirac spinor for the pseudospin doublet ($2g_{9/2}$,$1i_{11/2}$%
) is drawn in Fig.\ref{pseudowf} in several $\lambda $ values. The
wavefunction splitting of pseudospin doublet is obvious when $\lambda
/\lambda _{0}=1$. Different from the spin splitting, we can not see that the
pseudospin splitting reduces with $\lambda $ decreasing, which is consistent
with the case of level splitting.

In summary, Dirac Hamiltonian is scaled in the atomic units $\hbar =m=1$,
which allows us to take the non-relativistic limit by setting the speed of
light $c\rightarrow \infty $ or the Compton wavelength $\lambda \rightarrow
0 $. The evolution towards the non-relativistic limit is investigated from
the solutions of Dirac equation by a continuous transformation of the
parameter $\lambda $. The solutions of Dirac equation corresponding to $%
\lambda =\hbar /mc$ and $\lambda =0$ represent respectively the relativistic
result and that in non-relativistic limit. To transform the parameter $%
\lambda $ from $\hbar /mc$ to $0$, the solutions of Dirac equation show the
evolution from the relativistic to the non-relativistic limit. The
relativistic effects of the spin and pseudospin symmetries are checked from
the solutions of Dirac equation with the parameter $\lambda $. It shows the
spin splittings decrease monotonously with $\lambda $ reducing for all the
spin partners. When $\lambda $ is reduced to zero, the spin-orbit splittings
disappear, which is in agreement with the result in the non-relativistic
calculations. For the pseudospin symmetry, the energy splittings increase in
several partners, no change, or even decrease in another some partners.
Compared with the spin symmetry, the origin of pseudospin symmetry is more
complicated, and cannot be attributed uniquely to the relativistic effect.
The quantum number of single-particle states and the shape of potential make
important influence on this symmetry. By the perturbation calculations of
Dirac Hamiltonian, the various energy splitting behaviors of both spin and
pseudospin doublets with $\lambda $ are explained, which origins from the
different contributions of each component to the energy splittings. The
result supports the claim in Ref.\cite{Liang11}, the spin splitting can be
treated as a perturbation, while the pseudospin splitting can not be
regarded as a perturbation. The same conclusion can also be obtained from
the wavefunction.

This work was partly supported by the National Natural Science Foundation of
China under Grant No.11175001, the Excellent Talents Cultivation Foundation
of Anhui Province under Grant No.2007Z018, the Natural Science Foundation of
Anhui Province under Grant No.11040606M07, the Education Committee
Foundation of Anhui Province under Grant No. KJ2009A129, and the 211 Project
of Anhui University.


\begin{table}[h!]
\caption{The relativistic bound energies ($E=\protect\varepsilon-m$, in MeV)
of a Dirac particle for Woods-Saxon potential with $\protect\lambda/\protect%
\lambda_0=1,0.5,0.1,0.01,0.001$. The last column represents the
non-relativistic results, which are obtained from the solutions of the
Schr\"odinger equation $H\protect\psi (r)=E\protect\psi (r)$ with $H=-\frac{%
\hbar ^{2}}{2m}\left( \frac{d^{2}}{dr^{2}}-\frac{l(l+1)}{r^{2}}\right)
+\Sigma (r)$.}%
\begin{tabular}{c|ccccccc}
\hline\hline
$\lambda/\lambda_0$ & 1 & 0.5 & 0.1 & 0.01 & 0.001 & non &  \\ \hline
$1s_{1/2}$ & -59.206 & -60.756 & -61.109 & -61.123 & -61.123 & -61.123 &  \\
$2s_{1/2}$ & -41.592 & -47.056 & -48.305 & -48.353 & -48.353 & -48.353 &  \\
$3s_{1/2}$ & -18.358 & -27.968 & -30.287 & -30.376 & -30.377 & -30.377 &  \\
$1p_{3/2}$ & -52.763 & -55.631 & -56.306 & -56.332 & -56.332 & -56.332 &  \\
$1p_{1/2}$ & -52.263 & -55.578 & -56.304 & -56.332 & -56.332 &  &  \\
$2p_{3/2}$ & -31.401 & -38.680 & -40.396 & -40.462 & -40.463 & -40.463 &  \\
$2p_{1/2}$ & -30.611 & -38.576 & -40.393 & -40.462 & -40.463 &  &  \\
$3p_{3/2}$ & -7.694 & -17.957 & -20.633 & -20.737 & -20.738 & -20.738 &  \\
$3p_{1/2}$ & -6.999 & -17.822 & -20.628 & -20.737 & -20.738 &  &  \\
$1d_{5/2}$ & -45.234 & -49.459 & -50.493 & -50.533 & -50.534 & -50.534 &  \\
$1d_{3/2}$ & -44.055 & -49.329 & -50.489 & -50.533 & -50.534 &  &  \\
$2d_{5/2}$ & -20.999 & -29.752 & -31.897 & -31.980 & -31.981 & -31.981 &  \\
$2d_{3/2}$ & -19.573 & -29.543 & -31.890 & -31.980 & -31.981 &  &  \\
$1f_{7/2}$ & -36.882 & -42.381 & -43.786 & -43.841 & -43.842 & -43.842 &  \\
$1f_{5/2}$ & -34.775 & -42.137 & -43.779 & -43.841 & -43.842 &  &  \\
$2f_{7/2}$ & -10.759 & -20.436 & -22.933 & -23.031 & -23.032 & -23.031 &  \\
$2f_{5/2}$ & -8.777 & -20.102 & -22.922 & -23.031 & -23.032 &  &  \\
$1g_{9/2}$ & -27.921 & -34.508 & -36.276 & -36.346 & -36.347 & -36.346 &  \\
$1g_{7/2}$ & -24.701 & -34.114 & -36.264 & -36.346 & -36.347 &  &  \\
$1h_{11/2}$ & -18.545 & -25.944 & -28.044 & -28.127 & -28.128 & -28.128 &
\\
$1h_{9/2}$ & -14.117 & -25.366 & -28.025 & -28.127 & -28.128 &  &  \\
$1i_{13/2}$ & -8.942 & -16.792 & -19.167 & -19.262 & -19.263 & -19.263 &  \\
$1i_{11/2}$ & -3.361 & -16.000 & -19.141 & -19.262 & -19.263 &  &  \\
\hline\hline
\end{tabular}%
\end{table}

\begin{figure}[tbp]
\includegraphics[width=8cm]{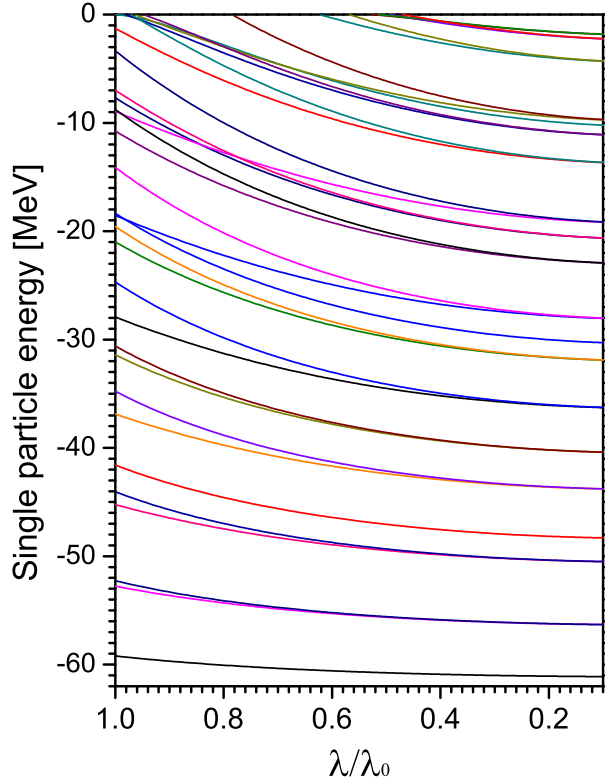}
\caption{(Color online) Variation of single particle energy with $\protect%
\lambda/\protect\lambda_0$.}
\end{figure}

\begin{figure}[tbp]
\includegraphics[width=8cm]{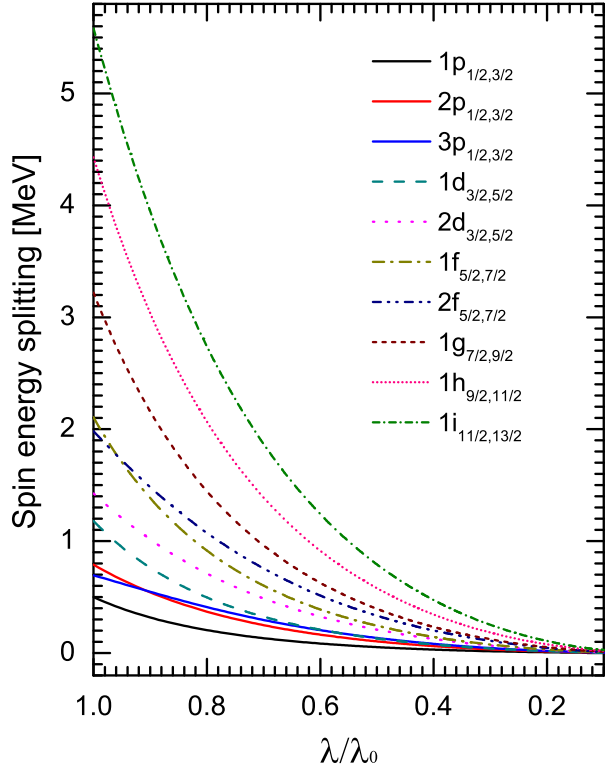}
\caption{(Color online) The energy splittings of spin doublets $\Delta
E=E_{n,l-1/2}-E_{n,l+1/2}$ varying with $\protect\lambda/\protect\lambda_0$.}
\end{figure}

\begin{figure}[tbp]
\includegraphics[width=8cm]{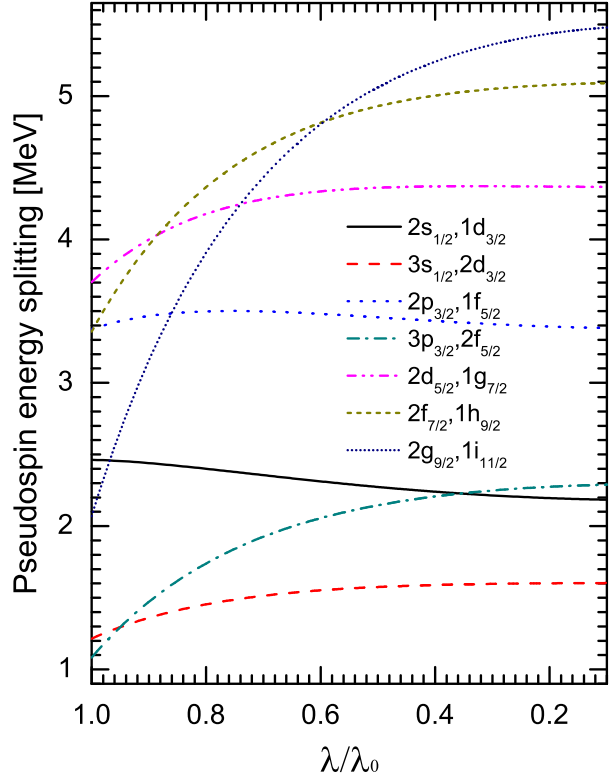}
\caption{(Color online) The energy splittings of pseudospin doublets $\Delta
E=E_{n,\tilde{l}-1/2}-E_{n-1,\tilde{l}+1/2}$ varying with $\protect\lambda/%
\protect\lambda_0$.}
\end{figure}

\begin{figure}[tbp]
\includegraphics[width=14.cm]{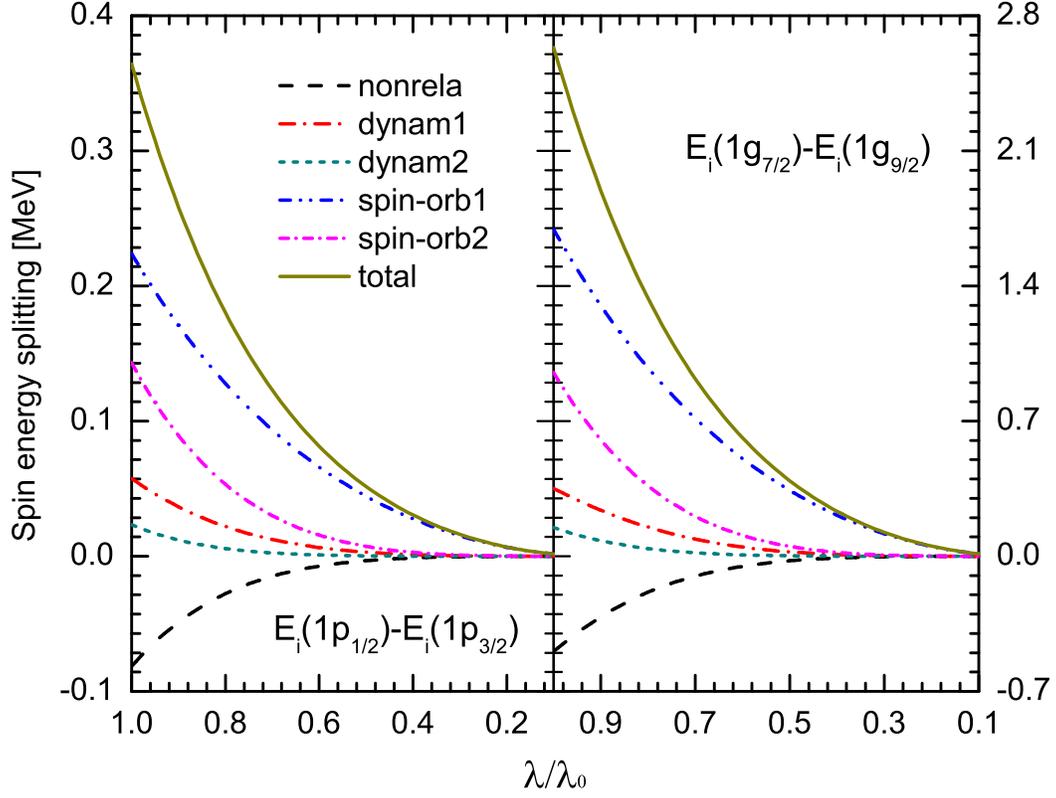}
\caption{(Color online) The spin energy splittings of each component $O_i$ ($%
i=1, 2, \dots, 5$) varying with $\protect\lambda$, where the splittings
caused by the $O_1$, $O_2$, $\dots$, $O_5$ are respectively labelled as
nonrela, dynam1, spin-orb1, dynam2, spin-orb2, and the total energy
splitting is labelled as total. }
\label{spinsplitting}
\end{figure}

\begin{figure}[tbp]
\includegraphics[width=14.cm]{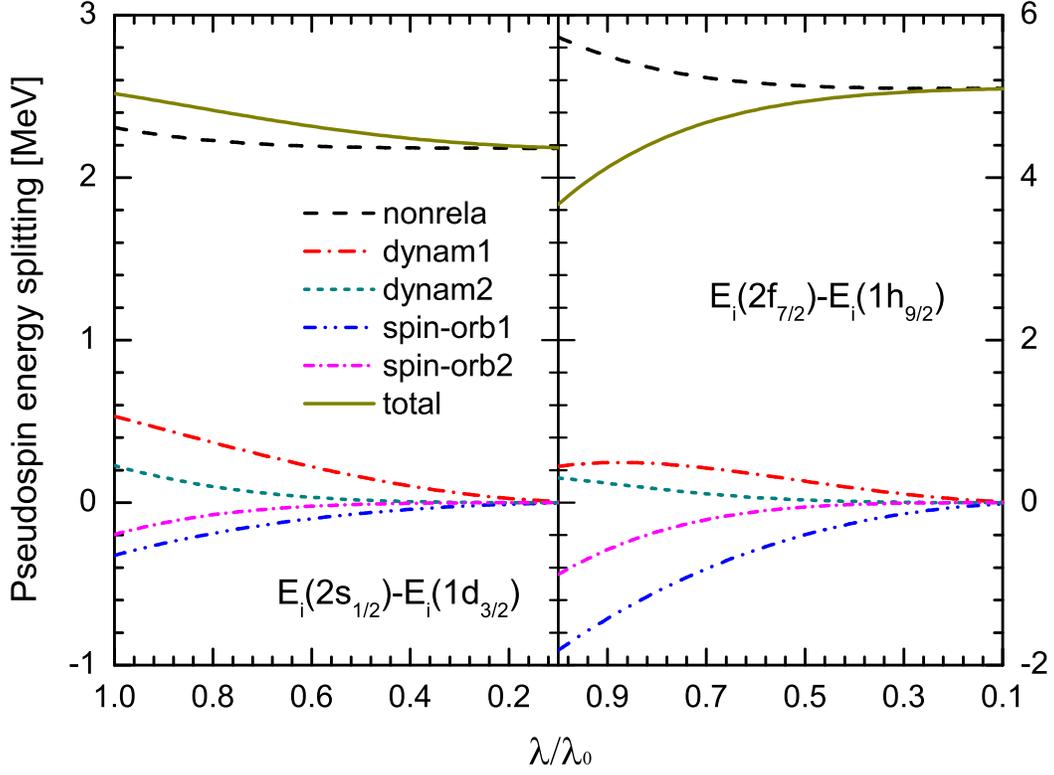}
\caption{(Color online) The same as Fig.\protect\ref{spinsplitting}, but for
the pseudospin energy splittings.}
\label{pseudosplitting}
\end{figure}

\begin{figure}[tbp]
\includegraphics[width=8.5cm]{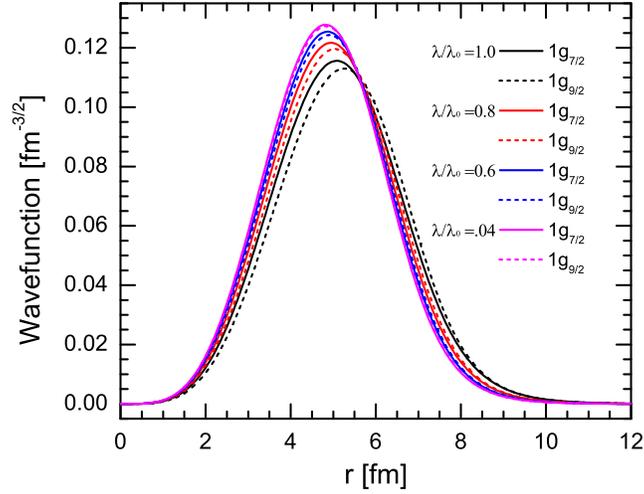}
\caption{(Color online) The upper component of Dirac spinor $G(r)/r$ for the
spin doublet ($1g_{7/2}$,$1g_{9/2}$) with $\protect\lambda/\protect\lambda%
_0=1.0,0.8,0.6,0.4$.}
\label{spinwf}
\end{figure}

\begin{figure}[tbp]
\includegraphics[width=8.5cm]{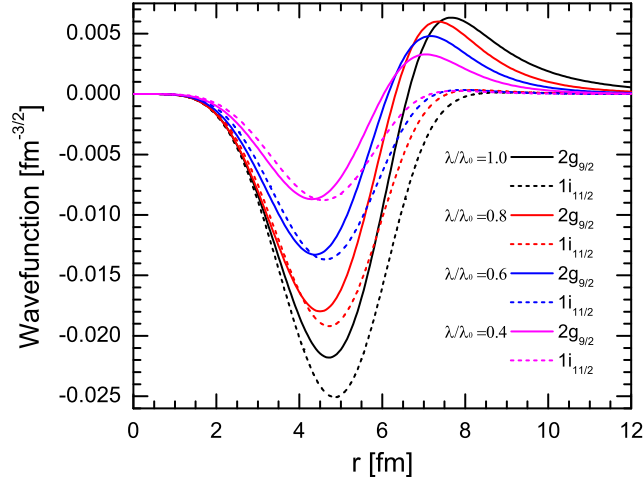}
\caption{(Color online) The lower component of Dirac spinor $F(r)/r$ for the
pseudospin doublet ($2g_{9/2}$,$1i_{11/2}$) with $\protect\lambda/\protect%
\lambda_0=1.0,0.8,0.6,0.4$.}
\label{pseudowf}
\end{figure}


\end{document}